\documentclass[pre,aps,onecolumn,superscriptaddress,notitlepage]{revtex4-2}

\usepackage[left=2cm,right=2cm,top=2cm,bottom=2cm]{geometry}
\usepackage{amsfonts}
\usepackage{amsmath}
\usepackage{amssymb}

\usepackage{bigints}
\usepackage{booktabs}

\usepackage{color}

\usepackage{dsfont}
\usepackage{float}
\usepackage{framed}
\usepackage{wrapfig}
\usepackage{tikz}

\usepackage{graphicx}
\usepackage{caption}
\usepackage[export]{adjustbox}

\usepackage{float}
\usepackage{indentfirst}
\usepackage{mathrsfs}
\usepackage{multirow}
\usepackage{setspace}
\usepackage{subdepth}
\usepackage{subfig}
\usepackage{titlesec}
\usepackage[dotinlabels]{titletoc}
\usepackage{wrapfig}
\usepackage[all]{xy}
\usepackage{young}
\usepackage[vcentermath]{youngtab}
\usepackage{relsize}
\usepackage{stackengine}
\usepackage{datetime}
\usepackage[linktocpage,colorlinks=true,allcolors=blue]{hyperref}

\usepackage{verbatim}

\newcommand{\be}{\begin{equation}}
\newcommand{\ee}{\end{equation}}\newcommand{\bml}{\begin{multline}}\newcommand{\emll}{\end{multline}}
\def\({\left(} \def\){\right)}\def\[{\left[} \def\]{\right]}
  \newcommand{\bea}{\begin{eqnarray}}\newcommand{\eea}{\end{eqnarray}}

\def\bf{\boldsymbol}

\usepackage{orcidlink}

\begin{document}

\title{How turbulent flows grow vorticity at a point}

\author{Timo Schorlepp\,\orcidlink{0000-0002-9143-8854}}
\email{timo.schorlepp@nyu.edu}
\affiliation{Courant Institute of Mathematical Sciences, New York University, New York, NY 10012, USA}

\author{Vladimir Rosenhaus\,\orcidlink{0000-0002-4906-6209}}
\email{vrosenhaus@gc.cuny.edu}
\affiliation{Initiative for the Theoretical Sciences, The Graduate Center, CUNY, New York, NY 10016, USA}

\author{Gregory Falkovich\,\orcidlink{0000-0002-0570-7895}}
\email{gregory.falkovich@weizmann.ac.il}
\affiliation{Weizmann Institute of Science, Rehovot 76100, Israel}

\date{Friday, 11$^{\rm th}$ September, 2026}


\begin{abstract}
We describe how strong vorticity fluctuations appear in three-dimensional incompressible turbulence driven by a large-scale random force.  By combining analytical and numerical methods, we construct a vorticity instanton: the most probable force history driving the flow to a prescribed large vorticity at a given point and time. We show how the smooth, large-scale force produces a long vortex filament with a thin, viscous-scale core. Our main finding is that the standard mechanism of amplifying a vortex tube through radial compression and axial stretching is only effective up to a threshold value.
Producing vorticity above the threshold requires an additional ingredient: vorticity waves propagating along the tube. The final vorticity is the result of two colliding waves. This mechanism has striking similarities to the one recently suggested by OpenAI for creating infinite vorticity in finite-time Navier--Stokes blow-up.
\end{abstract}


\maketitle


\begin{wrapfigure}{r}{6cm}
\centering
{%
\setlength{\fboxsep}{0pt}%
\setlength{\fboxrule}{1pt}%
\fbox{\includegraphics[width = .32 \textwidth]{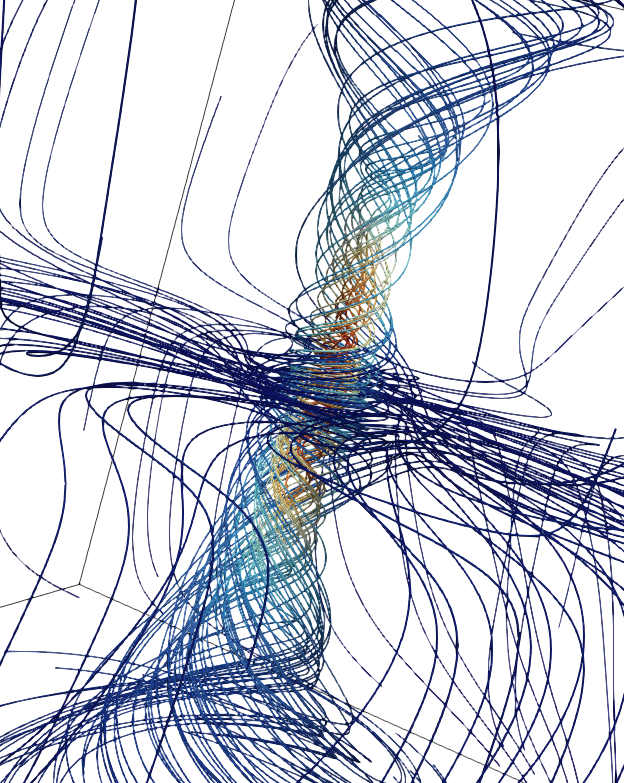}}%
}%
\caption{Instanton velocity field $\boldsymbol{u}$ at final time for $a = 1033$. Streamlines are colored according to the value of~$\omega_z$. }
\label{fig:3d-plot}
\end{wrapfigure}

\paragraph*{Introduction.} This note describes the most efficient route to generating large but finite vorticity, and thus has seemingly little to do with the Millennium problem of creating finite-time singularities in a viscous flow.  Yet, the underlying physical mechanism of colliding waves that we found and describe below is similar in spirit to the intricate ``spatially oscillating pulses'' utilized in Ref.~\cite{AI} and related works~\cite{Tristan,DMZZ} to generate unbounded vorticity growth. We elucidate the basic physics behind the waves: an interplay of the Coriolis force and radial squeezing.

\paragraph*{Setup.} 
We consider the incompressible three-dimensional Navier--Stokes equations driven by a Gaussian random force $\boldsymbol{f}$: 
\begin{equation}
    \partial_t \boldsymbol{u} + (\boldsymbol{u}\cdot\nabla)\boldsymbol{u} = -\nabla P + \Delta \boldsymbol{u} + \boldsymbol{f}\,, \quad
    \nabla\cdot \boldsymbol{u} = 0\,,
  \label{eq:nse}
\end{equation}
The random force is spatially smooth and temporally white:   $\langle \boldsymbol{f}(\boldsymbol{x},t)\boldsymbol{f}(\boldsymbol{0},0)\rangle=\delta(t)\chi(\boldsymbol{x})$, and $\chi$ is smooth. The probability to observe a given, large vorticity at the origin at the final time, $\omega_z({\bf 0},0)=a$, is $\mathbb{P}(\omega_z \approx a) \asymp \exp [-S(a)]$. Here, the action $S(a)$ is determined through the Gaussian force measure; the most probable way to generate vorticity corresponds to the minimal action. We call the optimal flow $\boldsymbol{u}(\boldsymbol{x}, t)$ an ``instanton''. Let us stress that the instanton is not an artificial construction: it is a physical flow configuration one naturally finds as the dominant vorticity structures in actual turbulent flows. The formalism to find it via optimization is described in the Methods section.

\begin{figure}
\centering
\includegraphics[width = .32 \textwidth]{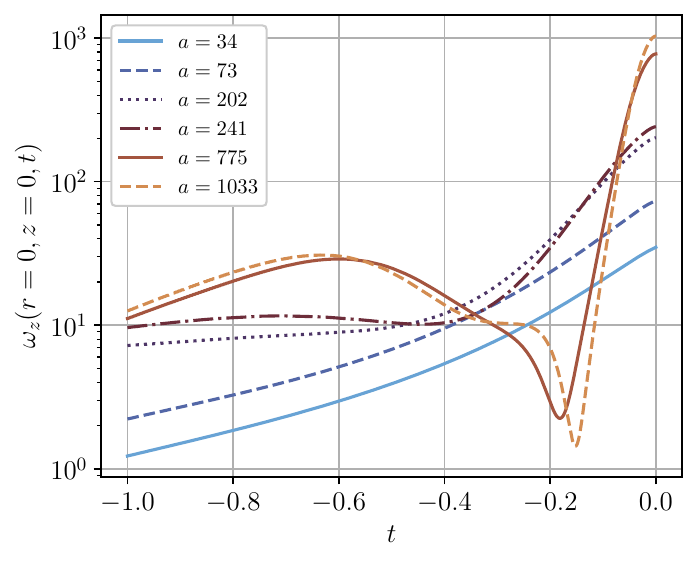}
\includegraphics[width = .32 \textwidth]{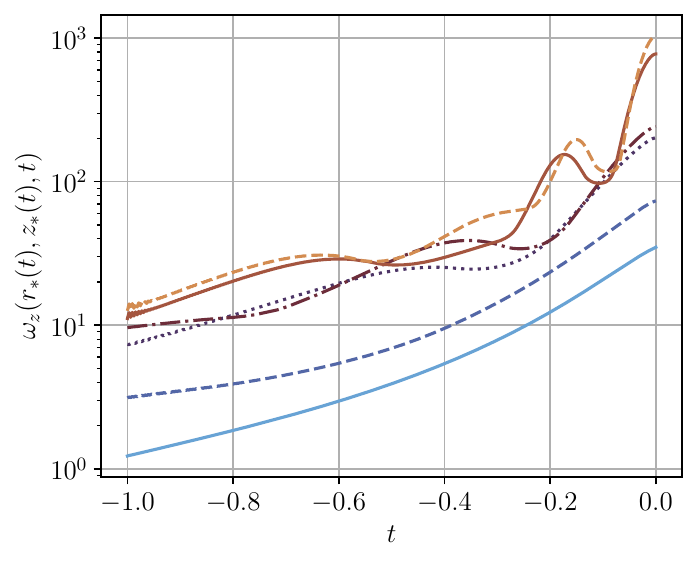}
\includegraphics[width = .32 \textwidth]{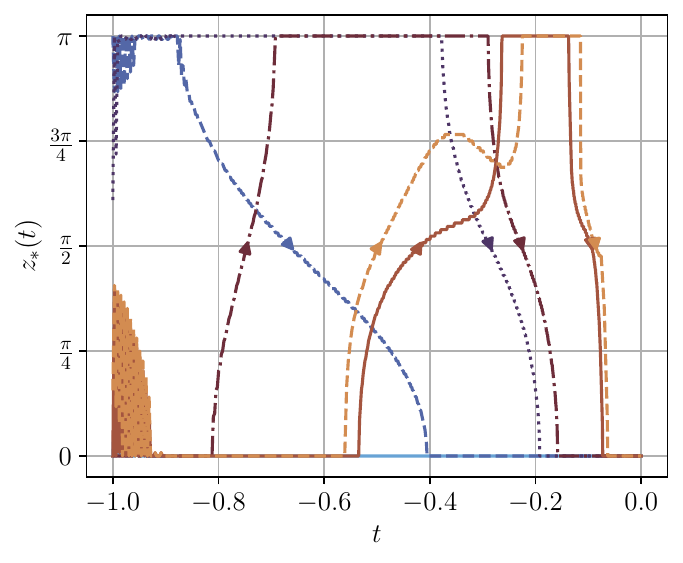}
\caption{Basic vorticity dynamics at different values of $a=\omega_z(r=0,z=0,t=0)$. Left: Vorticity evolution at the measurement point $r=z=0$. Center: Evolution of the maximal vorticity. Right: $z$-coordinate of the maximum.}
\label{fig:vort-growth-and-location}
\end{figure}
\paragraph*{Physical  mechanism.}  Figure~\ref{fig:3d-plot} shows the final-time flow configuration $\boldsymbol{u}$ generated by  numerically finding the most probable force $\boldsymbol{F}$ for a given $a$, assuming axial symmetry. Finding each such solution is computationally intensive (of course, not at the scale of OpenAI), taking several days of wall-clock time. We see a very thin and intense vortex filament in Fig.~\ref{fig:3d-plot}. How does a large-scale force manage to produce such a filament?

To generate large vorticity at a point (which we call the center), one can suggest the following strategy: 1) Apply~$F_\theta$ to make the fluid rotate around the axis. 2) Apply radial and axial forces to produce a large-scale flow directed radially inward in the central plane and outward from the center along the rotation axis. This will enhance the axial vorticity by squeezing the vortex tube. The radial strain is negative and axial strain is positive at the center. This strategy will grow vorticity at the center monotonically. Indeed,  such monotonic growth (using essentially a single physical mechanism for growth) was observed previously in all the known optimal fluctuations (instantons) in turbulence: for velocity and its gradients in the Burgers model  \cite{FKLM,Bal,CS,RG2}, for  the passive scalar \cite{BL}, for vorticity in the two-dimensional direct cascade \cite{FL2}, and for velocity in three-dimensional incompressible turbulence~\cite{FL1}.

\begin{figure}
\centering
\includegraphics[width = .32 \textwidth]{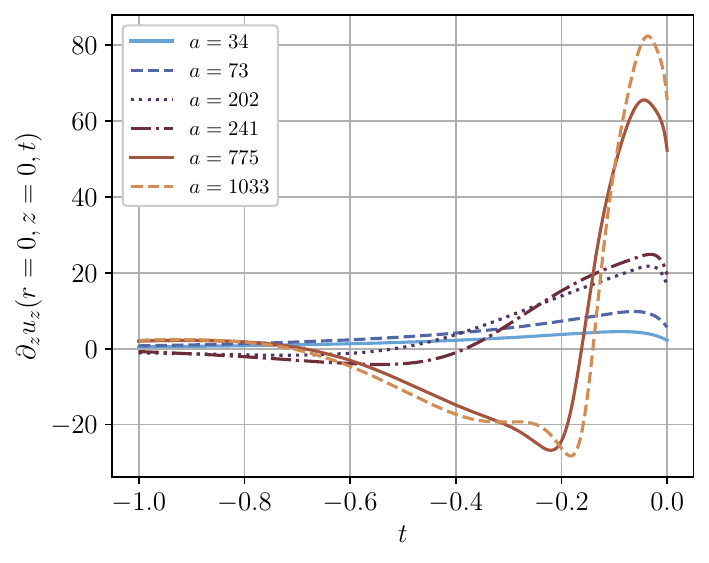}
\includegraphics[width = .32 \textwidth]{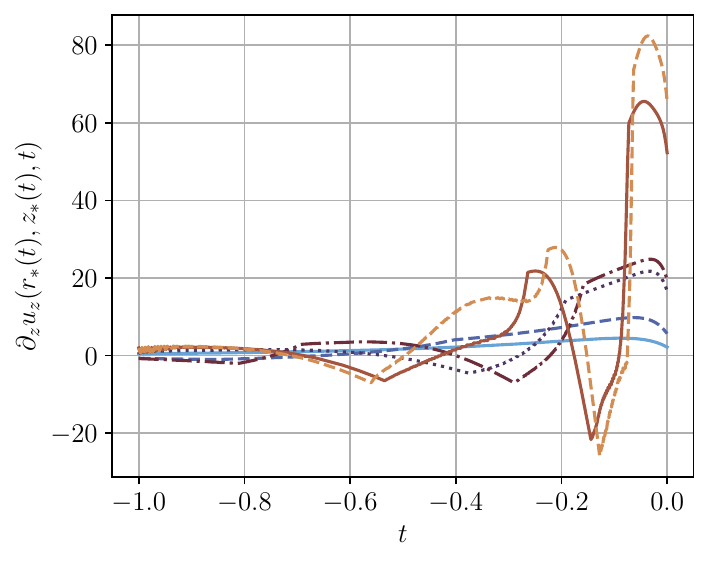}
\includegraphics[width = .32 \textwidth]{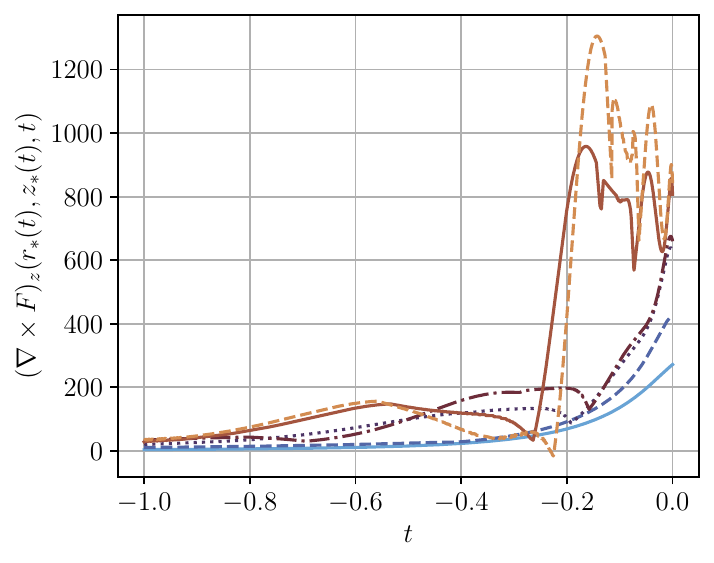}
\caption{Left: Strain evolution at the measurement point $r=z=0$. Center: Evolution of the strain at the location of maximal vorticity. Right: Forcing curl at the location of maximal vorticity.}
\label{fig:strain-and-force}
\end{figure}

\paragraph*{Coriolis force and necessity of waves.}  As is  already clear from the left panel of Fig.~\ref{fig:vort-growth-and-location}, simple monotonic growth at the measurement point only works for the low-vorticity values ($a<73$ for our choice of units and $\chi$). This is essentially a low-Re  regime in which the action grows quadratically with $a$: the probability is Gaussian,  $\log P(a)=-S\propto -a^2$. The vorticity at the center satisfies the equation $\partial_t\omega_z=\sigma\omega_z+\nu\Delta\omega_z+\partial F_\theta/\partial r$. For low $a$, the strain $\sigma=\partial_z u_z$ is positive at $z=0$ during the whole evolution (see Fig.~\ref{fig:strain-and-force}). For such low $a$, the vorticity and strain do not get too large too early, so  the Coriolis force does not prevent radial squeezing. As is known, the Coriolis force acts against an area change in the lateral plane of a rotating fluid, see, e.g., Ref.~\cite{Bat}.
 For the radial flow $u_r=-\sigma r$ in the region rotating with a background vorticity $\bar\omega$, we have the azimuthal Coriolis force $\bar\omega\sigma r$. Trying to keep the external force curl, $\partial F_\theta/\partial r$, larger than $\bar\omega\sigma$ is what makes the action grow quadratically with $a$. 
 
\begin{figure}
\centering
\includegraphics[width = .32 \textwidth]{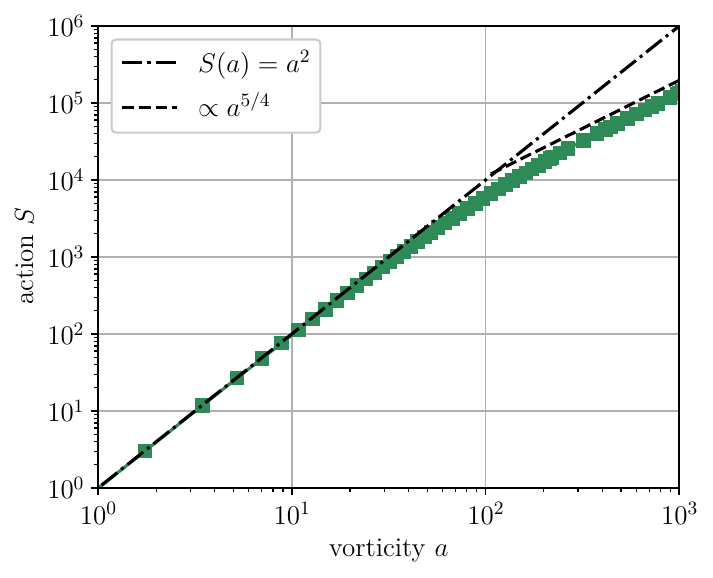}
\includegraphics[width = .32 \textwidth]{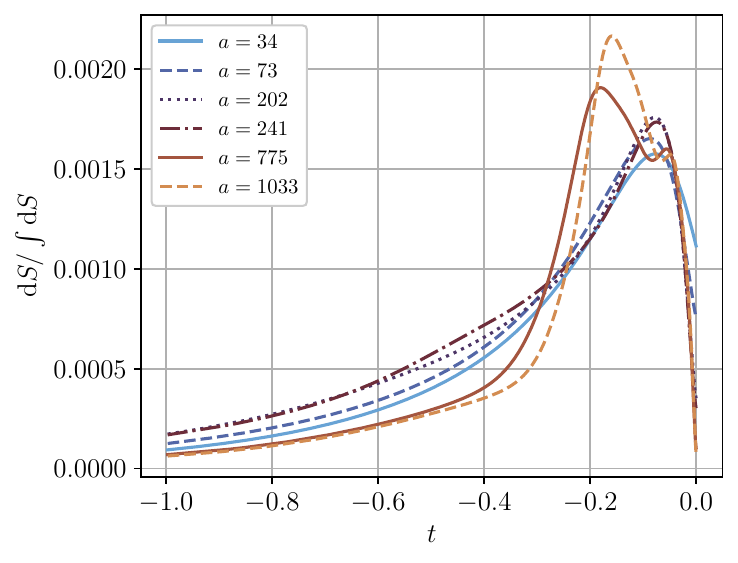}
\caption{Left: Instanton action $S$ vs.\ $a$. Right: corresponding action density over time}
\label{fig:action}
\end{figure}

 For larger values of $a$, the system finds a better (more probable) strategy in which the growth of $\boldsymbol{F}$  with $a$ is slower. This more optimal way to grow vorticity to higher values requires a more sophisticated strategy, which does not overcome the Coriolis force, but exploits it instead! Indeed, the data show that for $a=34$, we have $\partial F_\theta/\partial r\simeq 260>\bar\omega\sigma\simeq 35\cdot 3=105$, while for $a=73$ we have $\partial F_\theta/\partial r\simeq 420\simeq\bar\omega\sigma\simeq 75\cdot 7\simeq 525$. Also, the left panel of Fig.~\ref{fig:action} shows that the growth of $S(a)$ with $a$ is slower than quadratic for $a>80$. 

The question then is how to exploit the Coriolis force? It is a conservative restoring force, leading to a splitting of the vorticity maximum and propagation of vorticity waves along the rotation axis, as Fig.~\ref{fig:vort-growth-and-location} (right) demonstrates. The strategy taken by the optimal force is to grow vorticity inside these propagating pulses, making the large-scale pattern of $F_\theta$ follow them. To make the strategy more effective, the force $F_z$ generates an axial flow opposite to the propagation of the pulses in order to slow them down. Also, before every split, $\sigma$ is negative at the maximum, then becomes positive after the split, and of course $\sigma$ is positive during the final stage of growth at $z=0$ (Fig.~\ref{fig:strain-and-force}). The final large vorticity at the center is achieved by colliding two large pulses into the region of axial stretching and radial squeezing. The key point shared between the construction used in Ref.~\cite{AI} and our setting is that the vortex filament, whether it forms a true singularity in Ref.~\cite{AI} or merely a strong vortex tube at large $a$ here, has to be grown through the nonlinear equations of motion, and forcing has to act only indirectly, and not be the direct cause of vorticity growth in the ``naive'' way. Otherwise, in the blow-up case, a singularity is not possible to achieve with smooth forcing, and in our case, it would be ``too expensive'' in the action minimization sense (that is improbable).

\paragraph*{Overview of phases.} Let us now discuss in more detail the different regimes of pulse splitting and propagation (for different values of $a$), whose most salient features are shown in Fig.~\ref{fig:vort-growth-and-location}.
For $a \gtrsim 80$, the Coriolis force is large enough to provide splitting of the vorticity maximum. The optimal strategy for the next interval of $a$-values is to generate the vorticity maximum not at $z=0$, but at the opposite pole $z=\pi$ (as a reminder,   our $z$-axis is a ring $[0,2\pi]$, see Methods). Initially, the axial flow is compressing towards the vorticity maximum (that is, $\sigma<0$ there) to postpone the splitting and give more time for azimuthal acceleration. As seen in the right panel of Fig.~\ref{fig:vort-growth-and-location}, the center of any pulse propagates quickly at the beginning and at the end and slows down in the middle,  evidently because of a counterflow created by $F_z$. After the pole maximum splits into two counter-propagating pulses, $\sigma$ at the (moving) maximum turns positive in order to accelerate the vorticity growth by stretching. The pulses collide at $z=0$, where the desired value $\omega(0,0,0)=a$ is created. The pulses propagate faster for higher $a$. Splits appear later for instantons with larger $a$, probably because the confining flow (negative strain) is stronger. This strategy works for $73<a<202$. 

The first panel of Fig.~\ref{fig:action} shows the beginning of the new scaling of the action with $a$ around $a\simeq 200$. It corresponds to another change in strategy: to grow the final vorticity higher than $a\simeq200$, one needs more wave collisions.  The initial vorticity maximum is created and grows at $z=0$, then it splits into pulses which meet and fuse at $z=\pi$, then the second split creates the pulses which reunite at $z=0$, close to the moment of measurement.  

What we see as localized vorticity pulses propagating along the vortex tube can  alternatively be described in velocity terms as two fast-rotating rings. The two maxima of $\omega_z$ are on the axis (at $r=0$), while the round maxima of $u_\theta$ in the $z,r$-plane are at $r\simeq 1$ and moving along $z$, all the way until the short final time interval of order $\sigma^{-1}$. The ultimate fusion of velocity rings produces the fast-rotating tube with the velocity maximum spread over a thin long cylinder (as seen in the movie for $\boldsymbol{u}$ in the Supplementary Material).
To avoid misunderstanding, note that what we found is the collision of velocity rings (waves of $\omega_z$), not the much-studied collision of vortex rings (maxima of $\omega_\theta$). 

\begin{figure}
\centering
\includegraphics[width = .32 \textwidth]{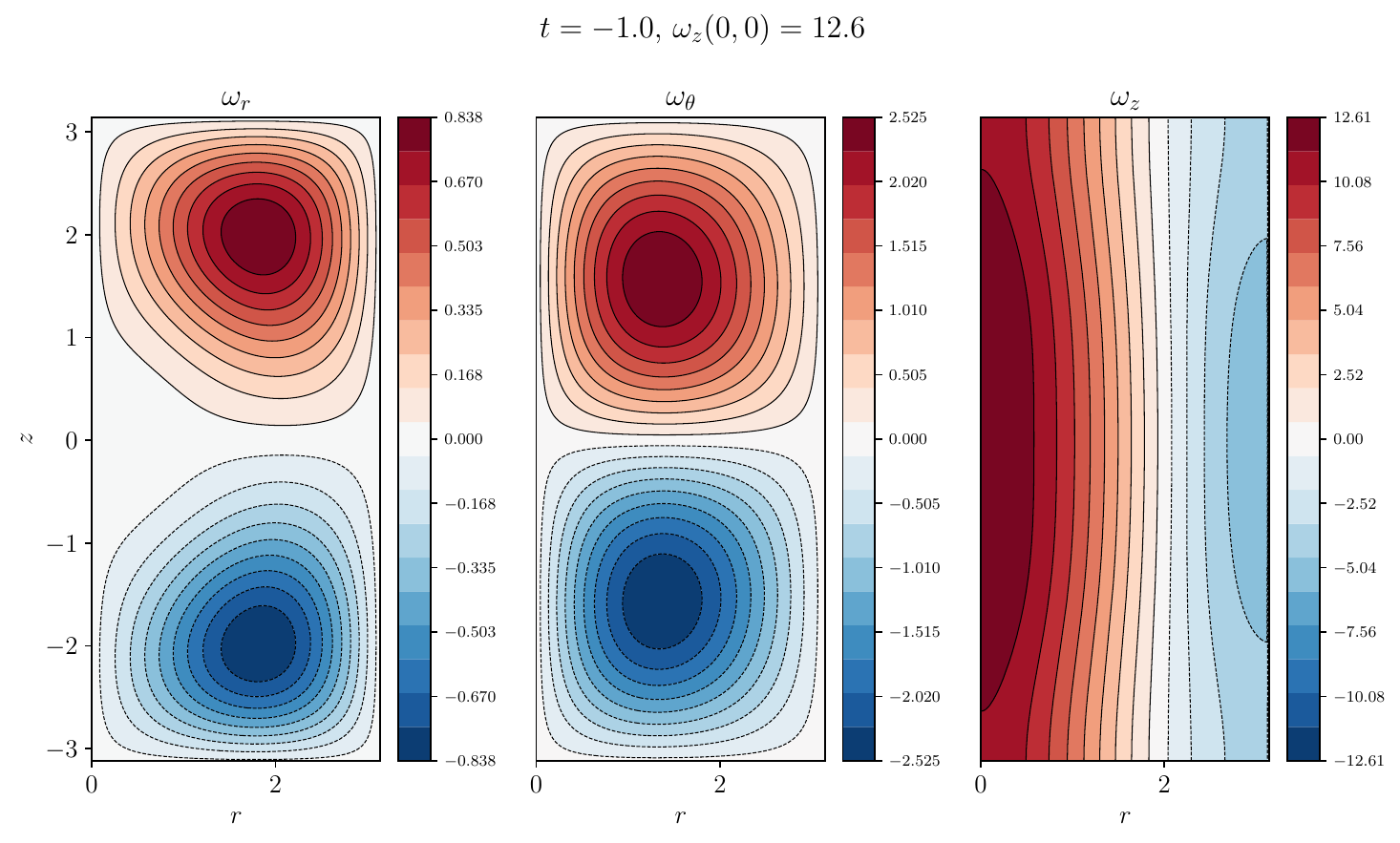}
\includegraphics[width = .32 \textwidth]{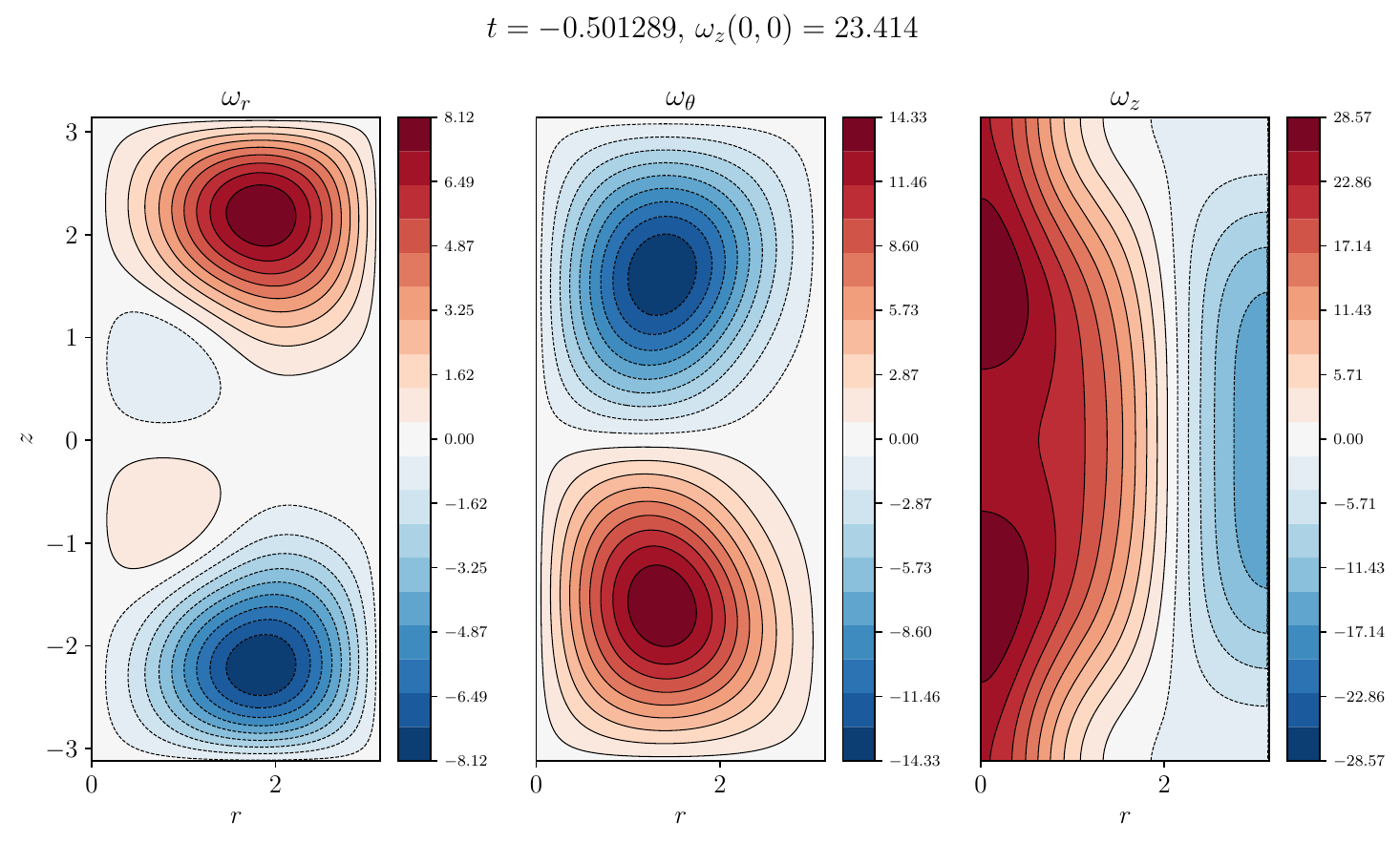}
\includegraphics[width = .32 \textwidth]{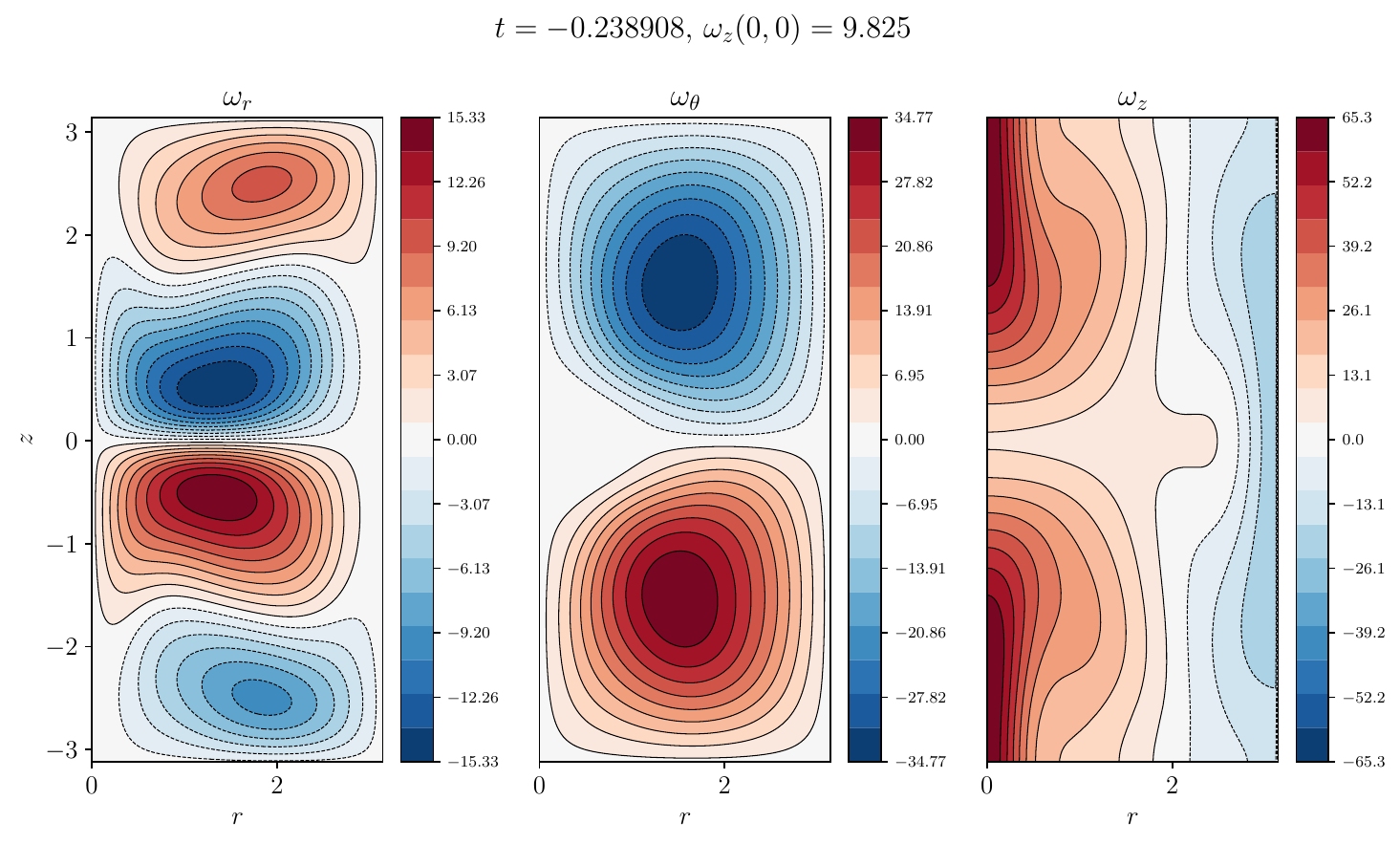}
\\
\includegraphics[width = .32 \textwidth]{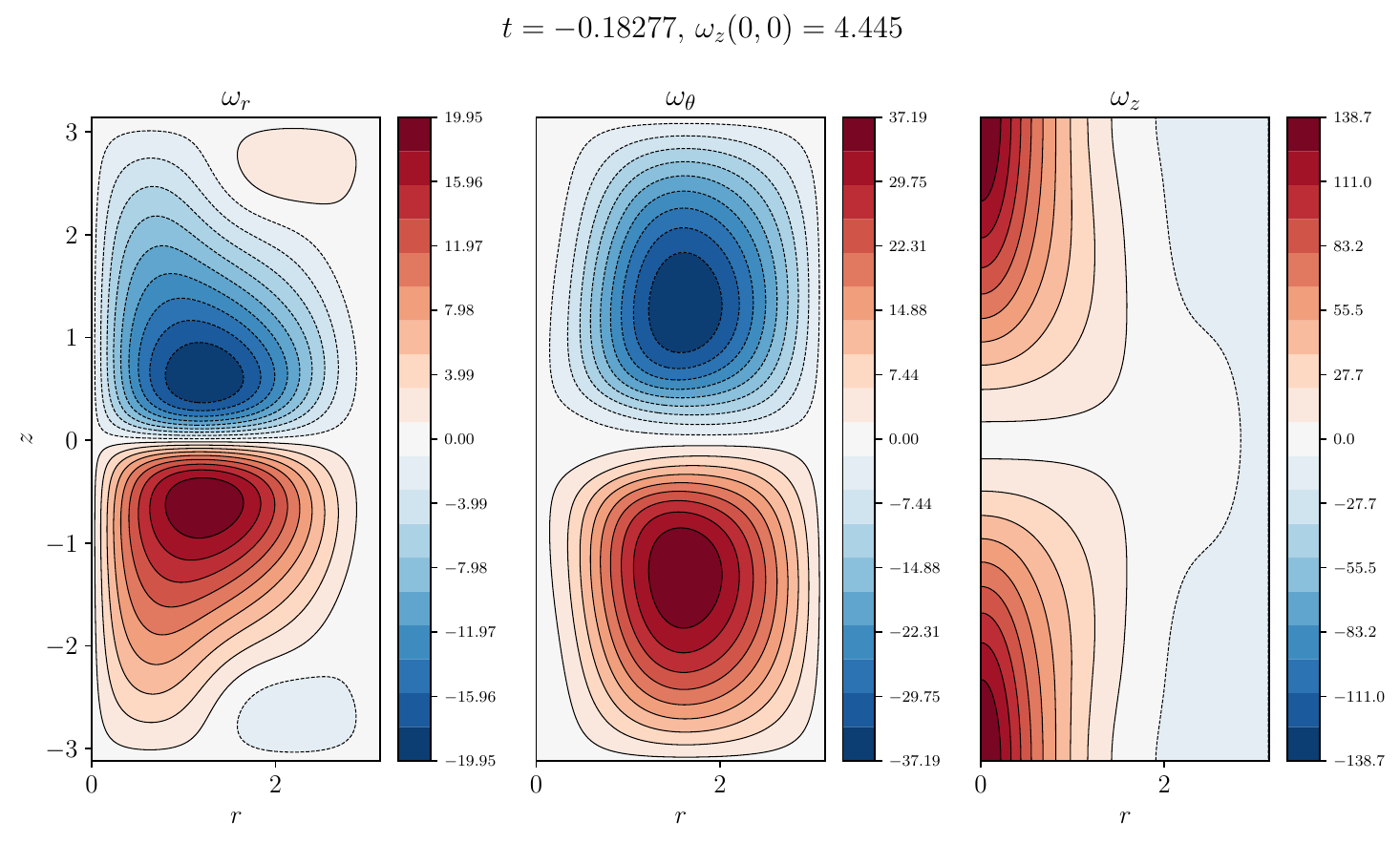}
\includegraphics[width = .32 \textwidth]{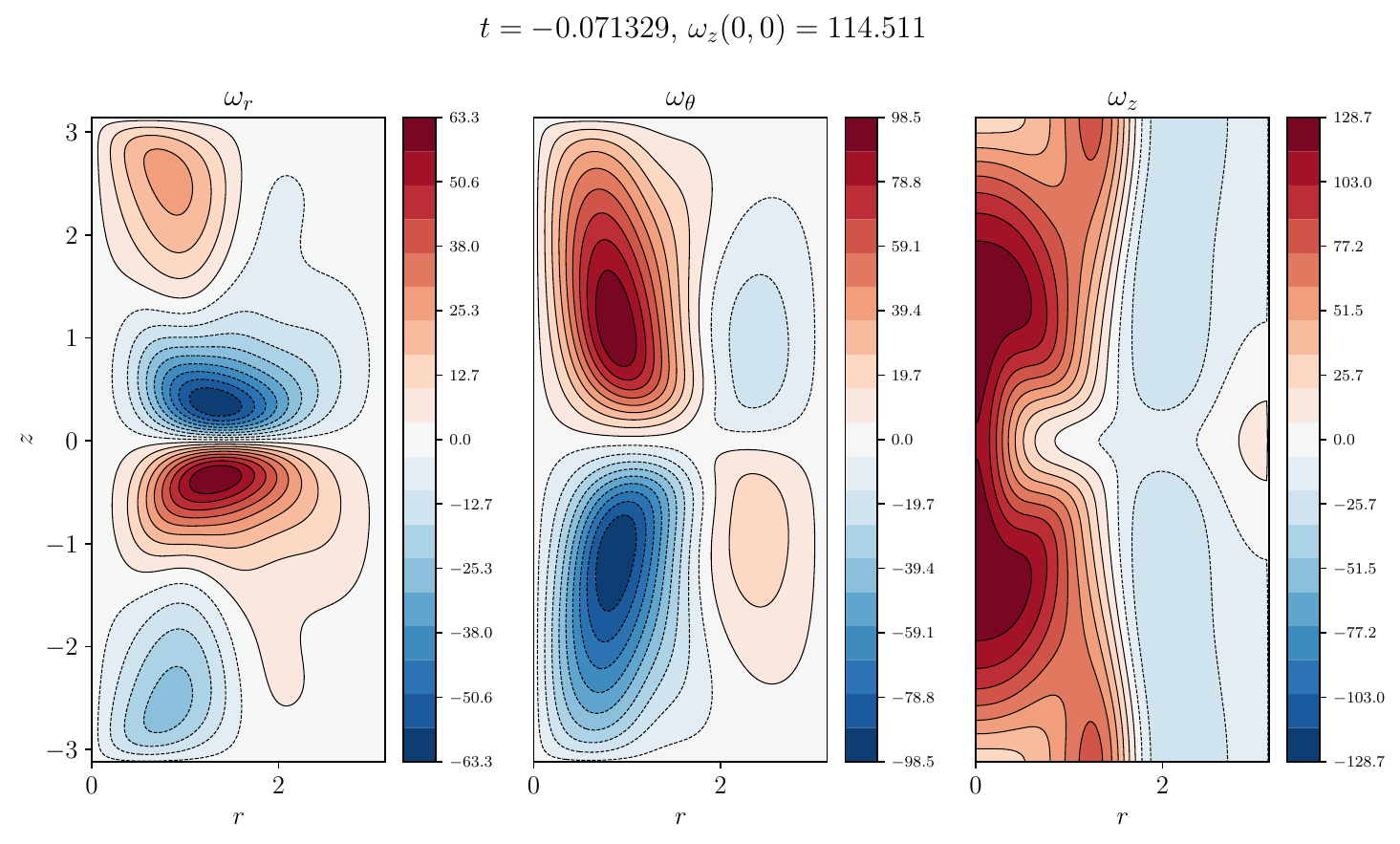}
\includegraphics[width = .32 \textwidth]{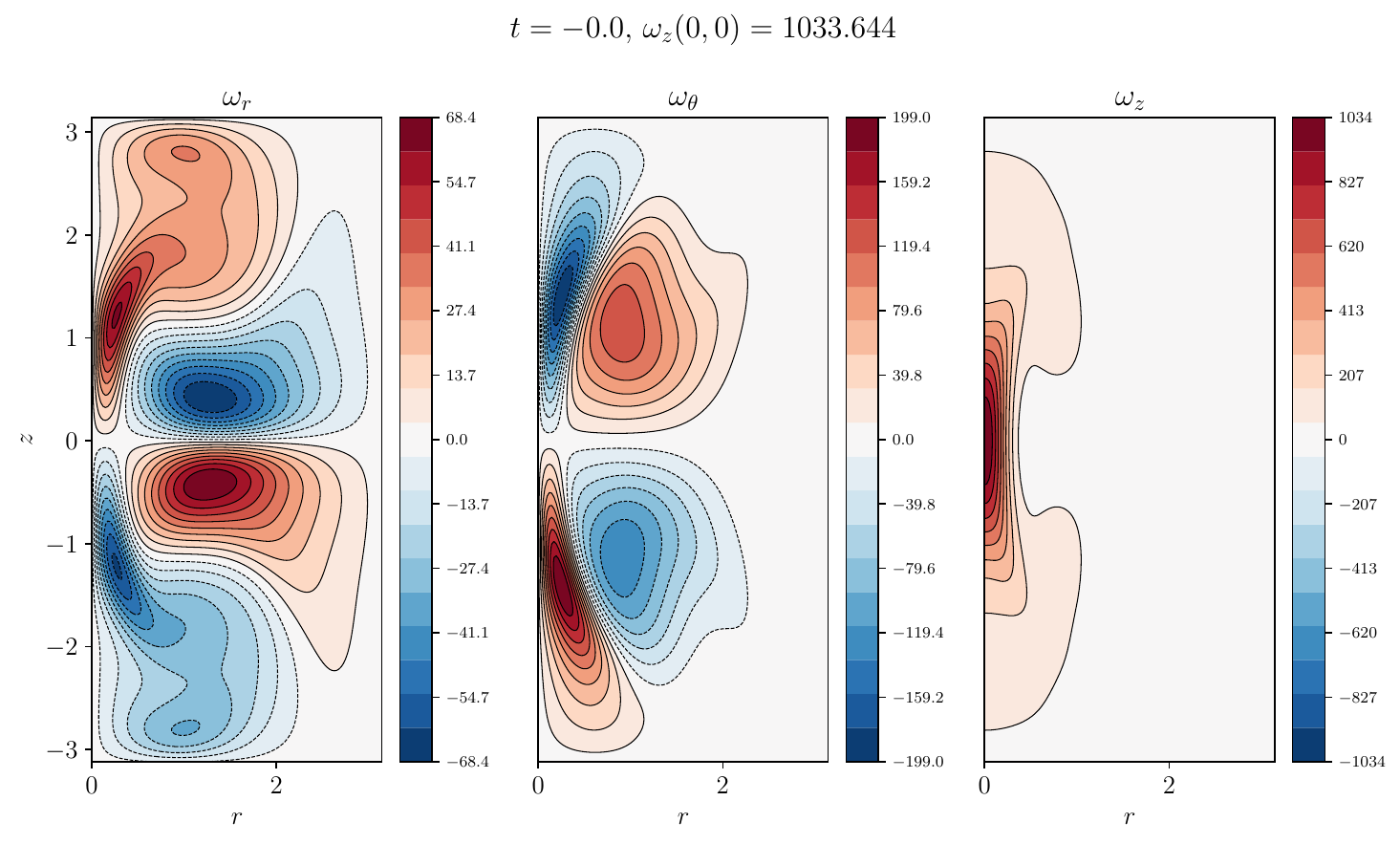}
\caption{Time evolution of vorticity components for $a = 1033$.
}
\label{fig:instanton}
\end{figure}

\paragraph*{Flow evolution at largest $a$.} Let us describe in some detail the history of the forces and the flow which brings the largest vorticity, $a=1033$, we have been able to obtain so far. The distinct stages are shown in Fig.~\ref{fig:instanton}. Complete movies of all  three components of the force, velocity and vorticity can be found in the Supplementary Material.

 The first stage is as for lower $a$: it involves rotation acceleration, radial squeezing and axial stretching of vorticity. This stage ends at $t\approx -0.75$ when 
the Coriolis force starts to be relevant.  
Then, the large-scale flow changes its geometry: At $t\approx -0.7$, the strain $\sigma$ at the center changes sign: the axial flow becomes directed towards the maximum at the center, trying to slow the split. The split starts anyway, at $t\approx-0.6$, when the vorticity at the center starts decreasing. During this transition from the first to the second stage, $-0.75<t<-0.6$, the azimuthal flow pattern $u_\theta(r,z)$ is the same, while the vertical-plane flow $u_r,u_z$ shifts by $\pi$ along $z$, as if preparing to squeeze vorticity into the opposite point $z=\pm\pi$. However, the $\omega_z$ maximum is still at $z=0$. 

The second stage then starts, with the central maximum splitting into two waves initially moving  with $c\simeq 40$, which is comparable with the estimate for the propagation speed, $\bar\omega\delta$,  where $\delta$ is the radius of the vortex tube. The large-scale pattern of $u_r$ and $u_z$ stays geometrically the same during the second stage. In particular, this shows that the pulses are not externally driven but propagate as (acoustic-like) perturbations of the tube. 
The pulses typically  have a triangular shape -- the leading edge is less steep. This is likely because the wave speed along the tube, $c=\bar \omega\delta$, decreases with the amplitude. As the waves run, vorticity inside continues to grow. This is visible in the velocity movie in the Supplementary Material, as we see two maxima of $u_\theta$ at $r\simeq 1$ move along $z$.  As seen from the force movie ($-0.25>t>-0.5$), the large-scale azimuthal force $F_\theta$ also splits into two maxima, each following the vorticity maxima all the way to the opposite side of the $z$-ring. As the maxima approach the pole at $z=\pm\pi$ (around $t\simeq-0.4$), they enter the radially squeezing flow, so that the vorticity growth is now both due to the growth of $u_\theta$ and squeezing of $\delta$. Around that time, the power of external forces and the action-generation rate $dS/dt$ start to grow sharply as shown in Fig.~\ref{fig:action} (right).  This is when the most work is done and the main cost in probability is paid. 

After the pulses fuse at the pole (around $t\simeq -0.19$), the new maximum has an even shorter lifetime since the vorticity inside is now larger. It promptly splits at $t\simeq -0.15$ into two counterpropagating waves. The power $dS/dt$ peaks around the time of the second split, when the largest force is $F_\theta$, see Fig.~\ref{fig:strain-and-force}. After the second split, the vertical-plane flow $u_r,u_z$ gradually shifts back to provide radial squeezing and axial stretching  at $z=0$. The second fusion of two vorticity pulses, now at $z=0$, happens around $t\simeq -0.06$, when the vortex tube around the center appears with $\omega_z(\boldsymbol{0},-0.06) \simeq 200$. The maximum of $\omega_z$ is already at $z=0$, while the two maxima of $u_\theta$ are still at $z\simeq\pm\pi/2$ and $r\simeq 1$. It is not clear at this stage if higher $a$ will lead to more than two splittings; this requires further study, as does observing such events in conditional averages from direct numerical simulation of Eq.~\eqref{eq:nse}~\cite{Bua2024}.

\begin{figure}
\begin{tikzpicture}
\node[anchor=south west, inner sep=0] (image1) at (0,0) {\includegraphics[width = \textwidth]{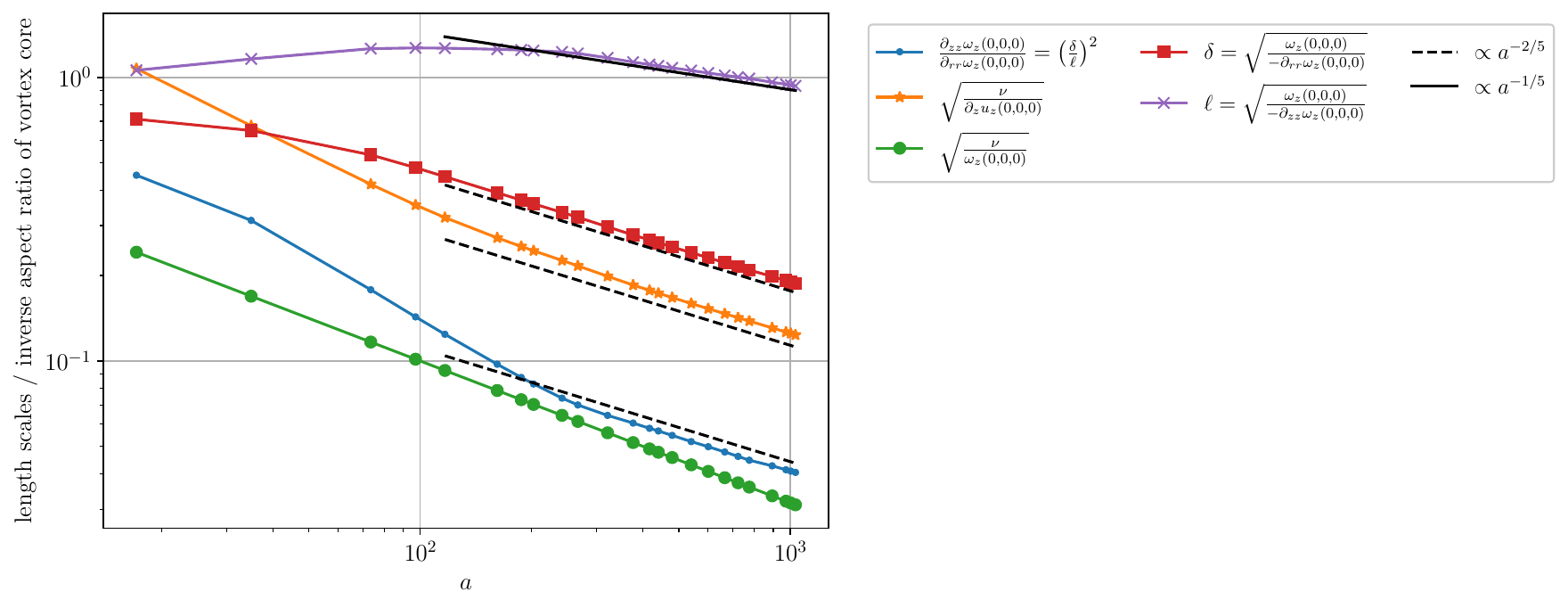}};
\node[anchor=south west, inner sep=0] (image2) at (10.5,0.1) {\includegraphics[width = .32 \textwidth]{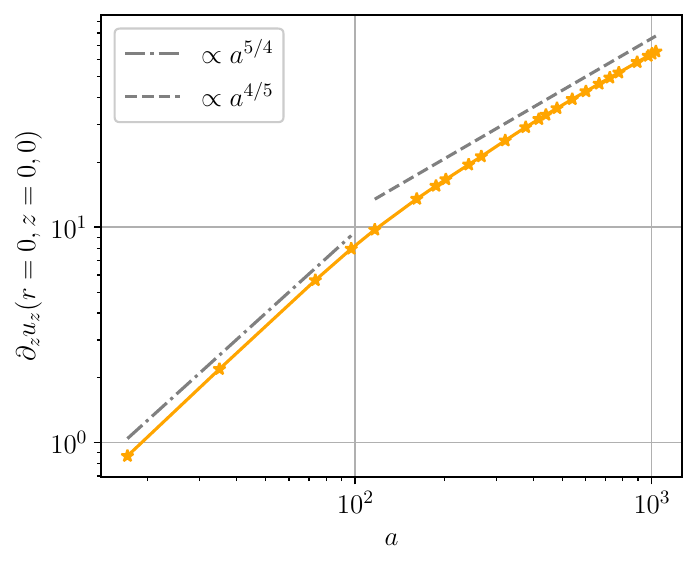}};
\end{tikzpicture}
\caption{Left: Dependence  of the final radius, length and aspect ratio of the vortex filament on the final vorticity. Right: Same for the strain at the final time at the origin.}
\label{fig:aspect}
\end{figure}

\paragraph*{The last stage and the scaling estimates.} The last stage ($t>-0.06$) of  vorticity growth costs no action (Fig.~\ref{fig:action}, right): it is force-free and is entirely driven by an interplay between pulse collisions and axial stretching, both of which increase the vorticity at the center.
Only around $t\simeq-0.02$ do the radial squeezing and axial stretching create a fast-rotating and very thin tube whose radial squeezing is balanced by viscous diffusion and axial stretching is balanced by compressing waves with  speed $c=a\delta$ (that is $\partial\omega/\partial t=\sigma\omega+c\partial\omega/\partial z$). This gives the estimates:
\begin{eqnarray}\delta\simeq\sqrt{\nu/\sigma}\,,\label{rad}\\ \sigma\simeq {c\over\ell}\simeq a{\delta\over\ell}\,.\label{ax}\end{eqnarray}
where $\ell$ is the size of the vortex core along the axis.
These two relations encode two projections of the velocity instanton equation. We need a third relation to express $\delta,\ell,\sigma$ via $a,\nu, \chi$. Mathematically, it must follow from the instanton equation for the field $\boldsymbol{p}$. We try to guess the right scaling, speculating that propagating waves must conserve  the vorticity volume integral, $M=\int \omega\,d \boldsymbol{r}$. If we measure $\omega$ in units of rms vorticity and measure $\delta,\ell$ in  box dimension units, then $M\simeq a \delta^2 \ell\simeq 1$. Substituting  Eqs.~(\ref{rad},\ref{ax}), we obtain
\begin{equation} \sigma\propto a^{4/5}\,,\quad \delta\propto a^{-2/5}\,,\quad \ell\propto a^{-1/5} \,.\label{fin}\end{equation}
These scaling laws agree with our numerics, see Fig.~\ref{fig:aspect}. Note, however, that  they are highly speculative and may change for even higher values of $a$. The main qualitative conclusions, though, seems robust (and agree with the ones made in Ref.~\cite{AI}): 1) The radial Reynolds number $\sigma\delta^2/\nu\simeq 1$ is independent of $a$, while the azimuthal Reynolds number $a\delta^2/\nu$ grows with $a$, and 2) The filament radius decreases with the vorticity faster than the axial length, so that the optimal configuration is indeed a thin long filament as often observed in turbulence experiments and numerics \cite{Sree, Gupta}. The final state is shown in the last three panels of Figure~\ref{fig:instanton}: a thin filament of $\omega_z$, and radially alternating signs of $\omega_\theta,\omega_r$.

One important conclusion for turbulence statistics is that the transverse velocity structure functions of all high orders scale as the distance squared for distances between $\delta$ and $\ell$. In other words, since the vortex filament is the most probable fluctuation, its codimension, 2,  is the lowest asymptotic anomalous exponent of the turbulent velocity field.

\paragraph*{Conclusion.}

The most important outcome of this note is the identification of waves as a necessary ingredient of vortex stretching: large-scale forces produce a vortex tube, radially compress, and stretch it  axially; sufficiently high vorticity maxima split into pulses whose propagation is slowed by the axial force. The azimuthal force follows these pulses, continuing to accelerate their spin. The ultimate peak in vorticity is attained when two pulses collide in the region of maximum stretching.

The need to slow down the pulses suggests an explanation for the remarkable phenomenon of axial symmetry breaking  discovered recently \cite{Timo}: for $a>85$,   a non-axially symmetric branch of instanton solution, with lower action, appears. Here, we find that waves emerge at a very similar threshold, $a \gtrsim 70$. This breaking of axial symmetry corresponds to the emergence of a helical vortex structure -- namely, a Kelvin wave with azimuthal wavenumber $m \neq 0$. Because the phase speed decreases with $m$, utilizing helical modes enables vorticity growth within a wave under weaker forcing. Reflection symmetry $z \to -z$ remains unbroken, strongly suggesting that these instantons also feature colliding waves.
\vspace{-\parskip}

 Since we are solving a statistical problem, the external force is an important ingredient that assigns probability measure to each flow realization. However, this does not mean  that our results are irrelevant for force-free (decaying) turbulence. The dynamical role of waves in creating large vorticity is likely to be universal -- even if the probability of such events differ --  with the ambient turbulent flow playing the role of the large-scale forcing considered here. 

\paragraph*{Acknowledgments.} This work did not make use of AI for the computations and writing.
GF and VR thank the Simons Center for Geometry and Physics for hospitality. The work of GF is supported by the grants 2024019 of BSF and 1204/24 of ISF and by the Excellence Center at WIS. The work of VR is supported in part by NSF grant PHY-2608451 and by BSF grant  2022113.

\paragraph*{Methods.}
Mathematically, the action can be written as $S(a) = \frac{1}{2}  \min_{\boldsymbol{\eta}}\lVert \boldsymbol{\eta} \rVert_{L^2}^2 $ s.t. $\omega_z = a$, where~$\omega_z$ depends on~$\boldsymbol{\eta}$ through Eq.~\eqref{eq:nse} with $\boldsymbol{f} = \chi^{1/2} * \boldsymbol{\eta}$. Here, $*$ denotes spatial convolution as defined below, and we define $\chi^{1/2}$ by $\chi^{1/2} * \chi^{1/2} = \chi$. In statistical physics, stochastic classical equations are described by a path integral, with an action in which a Lagrange multiplier --  physically a response field $\bf p$ --  enforces the equations of motion, 
\be
S = \int d t \int d\bf x\,  \bf p{\cdot} \big(  \partial_t \boldsymbol{u} + (\boldsymbol{u}\cdot\nabla)\boldsymbol{u} + \nabla P -\Delta \boldsymbol{u} -  \boldsymbol{f}\big)~,
\ee
where we omit the Lagrange multiplier enforcing $\nabla \cdot \boldsymbol{u} = 0$. Since the forcing is Gaussian we may integrate it out:
\begin{equation}
S = \int d t \int  d\bf x\,  \bf p{\cdot} \big(  \partial_t \boldsymbol{u} + (\boldsymbol{u}\cdot\nabla)\boldsymbol{u} + \nabla P -\Delta \boldsymbol{u} )- \frac{1}{2}\int dt \int d\bf x \int d\bf x'  p_i(\bf x,t) \chi_{ij}(\bf x{-}\bf x') p_j(\bf x',t)
\label{eq:action}
\end{equation}
Like in quantum mechanics, the superposition of many different flow histories $\bf u, \bf p$ contribute to realizing a particular vorticity at $t=0$. Since we are interested in extremely large vorticity at $t=0$, it is reasonable to hope that one history -- the saddle of the action -- is dominant. This saddle is referred to as an instanton, in rough analogy with the language in quantum field theory \cite{FKLM}. Extremizing the action~\eqref{eq:action} gives \cite{GGS, Timo}
\begin{align}\label{ie}
\begin{cases}
\partial_t \boldsymbol{u} + {\cal P} \left[(\boldsymbol{u}
\cdot\nabla) \boldsymbol{u} \right] - \Delta \boldsymbol{u} = \chi * \boldsymbol{p}\,,\\
\partial_t \boldsymbol{p} + {\cal P} \left[(\boldsymbol{u} \cdot\nabla) \boldsymbol{p} +
(\nabla \boldsymbol{p})^\top \boldsymbol{u}\right] + \Delta \boldsymbol{p} = 0\,,\\
\boldsymbol{u}(\cdot, -\infty) = 0\,, \quad \omega_z(\boldsymbol{0},0) = a\,, \quad \boldsymbol{p}(\boldsymbol{x}, 0) = \lambda \boldsymbol{e}_z \times \nabla \delta^{(3)}(\boldsymbol{x})
\end{cases}
\end{align}
where $\mathcal P_{ij}(\boldsymbol{k})=\delta_{ij}- k_i k_j / k^2$ projects onto the divergence-free part (a trick to circumvent the appearance of pressure), $\lambda$ is a Lagrange multiplier for the final time constraint $\omega_z(\boldsymbol{0},0) = a$, and $\delta^{(3)}$ the Dirac delta function, and on the right we have a convolution, $ \chi * \boldsymbol{p}(\bf x) = \int d \bf x' \chi(\bf x {-} \bf x') \boldsymbol{p}(\bf x')$. The first equation is Navier--Stokes with an effective forcing in terms of the response field $\bf p$, $\boldsymbol{F} = \chi * \boldsymbol{p}$, while $\bf p$ is itself self-consistently determined. We then have $S(a) = \tfrac12 \int dt \int d \boldsymbol{x} \; \boldsymbol{p} \cdot \chi * \boldsymbol{p}$ for the optimal value of the action, and $\boldsymbol{\eta} = \chi^{1/2} * \boldsymbol{p}$ above. The solutions of the extremum equations are called ``instantons'', since the high vorticity value appears only close to the final  instant of measurement ($t=0$).

Our setup and iterative numerical procedure to solve Eq.~\eqref{ie} is described in detail in Ref.~\cite{Timo}, where solutions were obtained for comparably low~$a$. Throughout this note, we assume axisymmetry of the flow, i.e., in cylindrical coordinates $(r,\theta,z)$ fields do not depend on $\theta$ (see discussion in the conclusion), as well as reflection symmetry about the $z = 0$ plane. The domain is a cylinder, with periodic boundary conditions in $z$. We use the same physical parameters and  axisymmetric code as in Ref.~\cite{Timo} (the latter, in turn, is based on Ref.~\cite{GS1991}). The only difference is that we do not set $\boldsymbol{u}(\cdot, -T) = \boldsymbol{0}$ (with $T = 1$), but instead determine it from linearized dynamics for $t \in (-\infty, -T]$. Hence, $\boldsymbol{u}(\cdot, -T)$ is part of the optimization. Oscillations at $t \simeq -T$ in some of the figures are due to the computational mode of the leapfrog method, which we dampen away as in Ref.~\cite{Timo}.

Our computations presume the force, box size and viscosity are of order unity, which is then true for the box-scale Reynolds number as well. 
Our results can be rescaled to arbitrary $Re$ by dividing $a$ by $Re^b$. To find $b$, note that the effective force in Eq.~\eqref{ie} scales as $\chi p$,  the Reynolds number scales as $\chi^{1/3}$, and $p$ scales as $S(a)/a\propto a^c$. Therefore, $b=3/c$. For our interval of $a$, we find $c\approx 1/4$ so that $b\approx12$.

\end{document}